\documentclass{jpp}
\usepackage{graphicx}
\usepackage{epstopdf, epsfig}
\usepackage{hyperref}

\usepackage{graphicx}
\usepackage{mathrsfs}
\usepackage{color}
\usepackage[table]{xcolor}
\usepackage{hyperref}
\hypersetup{colorlinks, citecolor=black, filecolor=black, linkcolor=black, urlcolor=black}

\expandafter\let\csname equation*\endcsname\relax
\expandafter\let\csname endequation*\endcsname\relax
\usepackage{amsmath}

\usepackage{dcolumn}
\usepackage{bm}

\definecolor{grayish}{RGB}{230,230,230}

\newcommand{\refEq}[1] {(\ref{#1})}

\newcommand{\romanNum}[1]{\uppercase\expandafter{\romannumeral#1}}

\shorttitle{Eliminating self-interaction through the parallel boundary}
\shortauthor{J. Ball, S. Brunner, and Ajay C.J.}

\title{Eliminating turbulent self-interaction through the parallel boundary condition in local gyrokinetic simulations}

\author{Justin Ball, Stephan Brunner, and Ajay C.J.}

\affiliation{Ecole Polytechnique F\'{e}d\'{e}rale de Lausanne (EPFL), Swiss Plasma Center (SPC), CH-1015 Lausanne, Switzerland}

\begin{document}

\maketitle

\begin{abstract}
In this work, we highlight an issue that may reduce the accuracy of many local nonlinear gyrokinetic simulations --- turbulent self-interaction through the parallel boundary condition. Given a sufficiently long parallel correlation length, individual turbulent eddies can span the full domain and ``bite their own tails,'' thereby altering their statistical properties. Such self-interaction is only modeled accurately when the simulation domain corresponds to a full flux surface, otherwise it is artificially strong. For Cyclone Base Case parameters and typical domain sizes, we find that this mechanism modifies the heat flux by roughly 40\% and it can be even more important. The effect is largest when using kinetic electrons, low magnetic shear, and strong turbulence drive (i.e. steep background gradients). It is found that parallel self-interaction can be eliminated by increasing the parallel length and/or the binormal width of the simulation domain until convergence is achieved.
\end{abstract}

\section{Introduction}
\label{sec:intro}

Local nonlinear gyrokinetic simulations are one of the most commonly used tools to assess turbulent transport in tokamaks. They solve the gyrokinetic model \citep{CattoLinearizedGyrokinetics1978, FriemanNonlinearGyrokinetics1982, BrizardGKreview2007, AbelGyrokineticsDeriv2012}, a nonlinear system of integro-differential equations that have been derived to model plasma turbulence as accurately as possible. Indeed, the primary approximation of gyrokinetics, $\rho_{\ast} \equiv \rho_{i} / a \ll 1$ (i.e. the tokamak minor radius $a$ is much larger than the ion gyroradius $\rho_{i}$), is satisfied in many existing tokamaks by factors of several hundred. Unfortunately, even with its approximations, such a high-fidelity model is very computationally expensive. For this reason, it has been crucial to streamline the calculations as much as possible.

\begin{figure}
	\centering
	\includegraphics[width=0.5\textwidth]{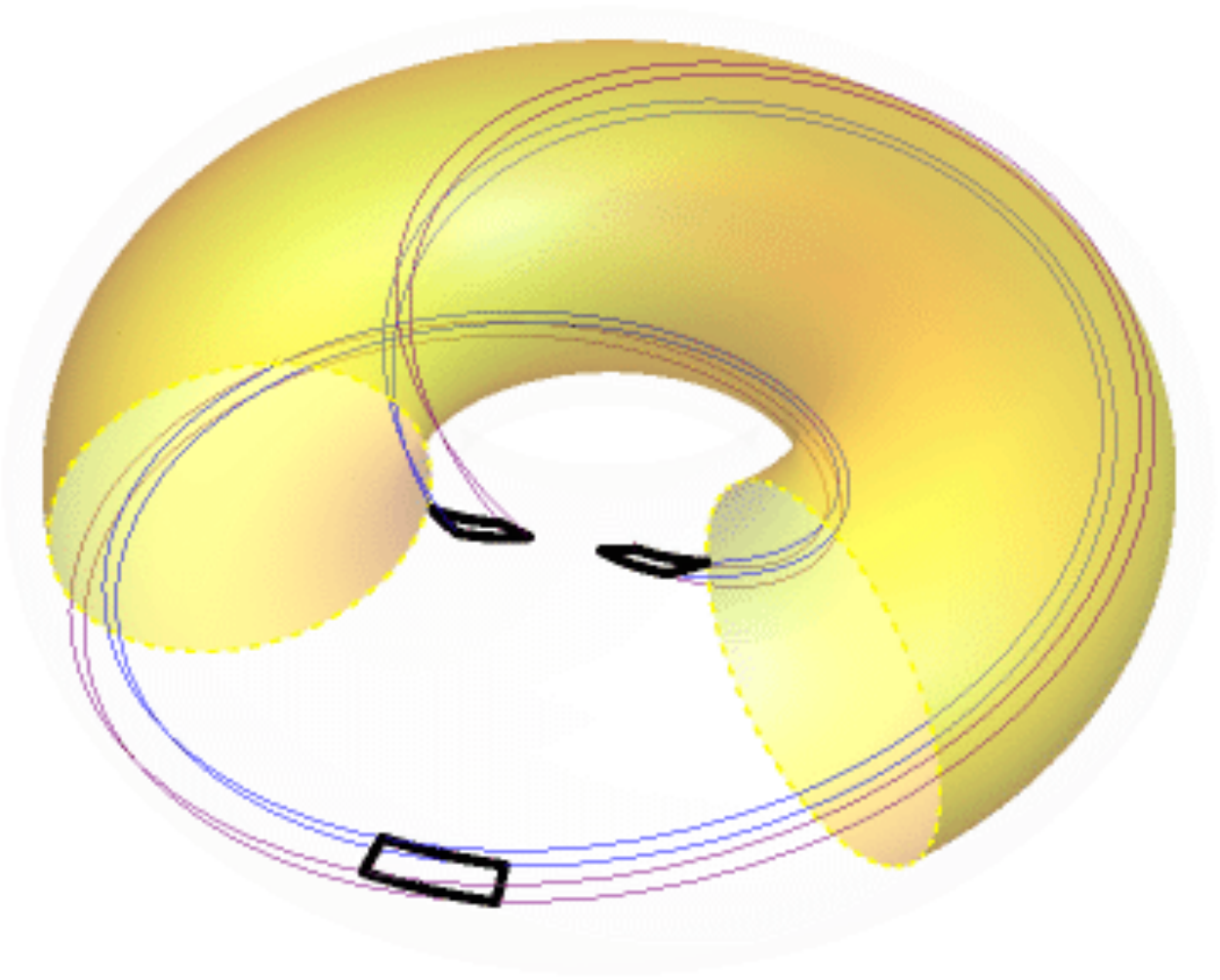}
	\caption{An example flux tube (thin blue and purple lines) that is one poloidal turn long with its cross-sectional shape indicated on the outboard midplane (thick black rectangle) and at both locations on the inboard midplane (thick black parallelograms). Also shown is the central flux surface of the flux tube (transparent yellow) with a toroidal wedge removed for visual clarity.}
	\label{fig:fluxTubeGeo}
\end{figure}

To this end, it is helpful to minimize the volume of the simulation domain and only include the minimal number of turbulent eddies needed to obtain a statistically-relevant representation of turbulence. Instead of modeling full magnetic flux surfaces across most of the plasma minor radius (referred to as ``global'' simulations), a calculation can get by with a much smaller domain (referred to as ``local'' simulations). Since turbulence in tokamaks is aligned with the magnetic field and is very anisotropic, it is important for the domain of a local simulation to have similar characteristics. Thus, the domain is long in the direction parallel to the magnetic field and quite narrow in the two perpendicular directions. Full flux surfaces are {\it not} usually modeled, nor is a large fraction of the minor radius. Typically, the domain has a rectangular cross-section on the outboard midplane and is deformed into a parallelogram at other poloidal locations due to the effect of magnetic shear. Such a domain is called a ``flux tube'' \citep{BeerBallooingCoordinates1995} (see figure \ref{fig:fluxTubeGeo}). It can enable the computational cost of local simulations to be orders of magnitude lower than global simulations, particularly when modeling large devices.

An important difference between local and global simulations concerns boundary conditions. In global simulations, the boundaries within the flux surface are straightforward --- because entire flux surfaces are modeled, the physical periodicity in the toroidal and poloidal directions can be implemented directly. However, the radial boundary condition for global simulations is less obvious. Typically Dirichlet boundary conditions are applied together with ``buffer regions,'' which are radial regions that contain sources of particles and energy \citep{CandyGYRO2003, GoerlerGlobalGENE2011} \footnote{Additionally, an artificial Krook damping operator is often used to prevent profile relaxation in the simulation region.}. Hence, it is important to monitor global simulations to ensure that the results are not sensitive to the particularities of this complex and artificial boundary condition.

In contrast, all of the boundary conditions for local simulations are elegant \citep{BeerBallooingCoordinates1995}, though they still require care (as is emphasized by this work). Since the domain of local simulations is narrow in the radial direction, the background plasma parameters (e.g. density, temperature) are well approximated by a constant value with a linear gradient to drive turbulence. This naturally reflects the inherent scale separation between fluctuations and background gradients, which is a fundamental assumption of the gyrokinetic model \citep{AbelGyrokineticsDeriv2012}. One consequence is that the turbulence experiences the same drive and background plasma conditions at opposing sides of the domain, so it should be {\it statistically} identical. This is also true in the binormal direction (i.e. the direction within the flux surface and perpendicular to the magnetic field). For simplicity, instead of generating {\it statistically} identical turbulence for these boundaries, it is substituted with {\it exactly} identical turbulence by using periodic boundary conditions. Using exactly identical turbulence instead of statistically identical turbulence does have the potential to introduce unphysical correlations into the system. However, as long as the correlation length of the turbulence is much smaller than the domain, this will not occur. In practice, the validity of this can be determined by testing for convergence in the widths (radial and binormal) of the domain.

For nonlinear simulations, the boundary condition in the parallel direction is similar, but with one additional complicating factor --- the effect of magnetic shear. In this work we will study the parallel boundary condition and its effect on convergence in the parallel and binormal sizes of the simulation domain. We will find that the domain length typically used by the community, one poloidal turn, can introduce unphysical turbulent correlations, directly affecting the accuracy of the results. Individual turbulent eddies can remain correlated across the entire parallel length of the domain and interact with themselves, which is unphysical unless the domain corresponds to the full flux surface. This was first discussed in the original flux tube model paper \citep{BeerBallooingCoordinates1995} in the context of gyrofluid simulations with adiabatic electrons. Subsequent gyrokinetic studies \citep{CandyShearLayers2005, WaltzShearLayers2006, Dominski15} investigated the ability of this mechanism to generate localized, steady corrugations in the background profiles. More recent work considered the self-interaction of linear eigenmodes and revealed that their parallel self-interaction can even interfere with the numerical convergence of nonlinear simulations with respect to the domain size \citep{Weikl2018, AjaySelfInteraction2020} \footnote{Note that the self-interaction considered in these previous works (i.e. the nonlinear interaction of a {\it linear eigenmode} with itself) is closely related to the self-interaction considered here (i.e. the interaction of a {\it nonlinear turbulent eddy} with itself), but the two are distinct. Most notably, in the absence of magnetic shear, linear eigenmodes would not exhibit self-interaction, but nonlinear eddies still could.}. Additionally, it has been explored in gyrokinetic simulations of low magnetic shear stellarators \citep{MartinParallelBCstell2018, FaberSelfInteractionStellarator2018}. {\it This paper will show that parallel self-interaction can be eliminated by lengthening the simulation domain in the parallel direction to multiple poloidal turns or making the binormal width of the domain large.} It is the first work to demonstrate that these two methods produce the same result and presents a novel, simple test for the presence of parallel self-interaction (i.e. the radial dependence of the parallel correlation function shown in figure \ref{fig:parCorr}).

\section{The parallel boundary condition}
\label{sec:parallelBC}

The parallel boundary condition generally used by nonlinear flux tube simulations is called the ``twist-and-shift'' condition \citep{BeerBallooingCoordinates1995}. As with the perpendicular boundaries, it requires the two ends of the flux tube to have statistically identical turbulence. Finding statistically identical turbulence is non-trivial because toroidicity and plasma shaping mean that the statistical properties of turbulence change with poloidal angle. Thus, to be statistically identical the two ends of the flux tube must be at the same {\it poloidal} angle $\chi$, though, due to axisymmetry, they are still free to be at different {\it toroidal} angles  $\zeta$. This constraint requires the length of the domain to be an integer number of poloidal circuits, which we represent by $N_{\text{pol}}$.

\begin{figure}
	\centering
	\includegraphics[width=0.8\textwidth]{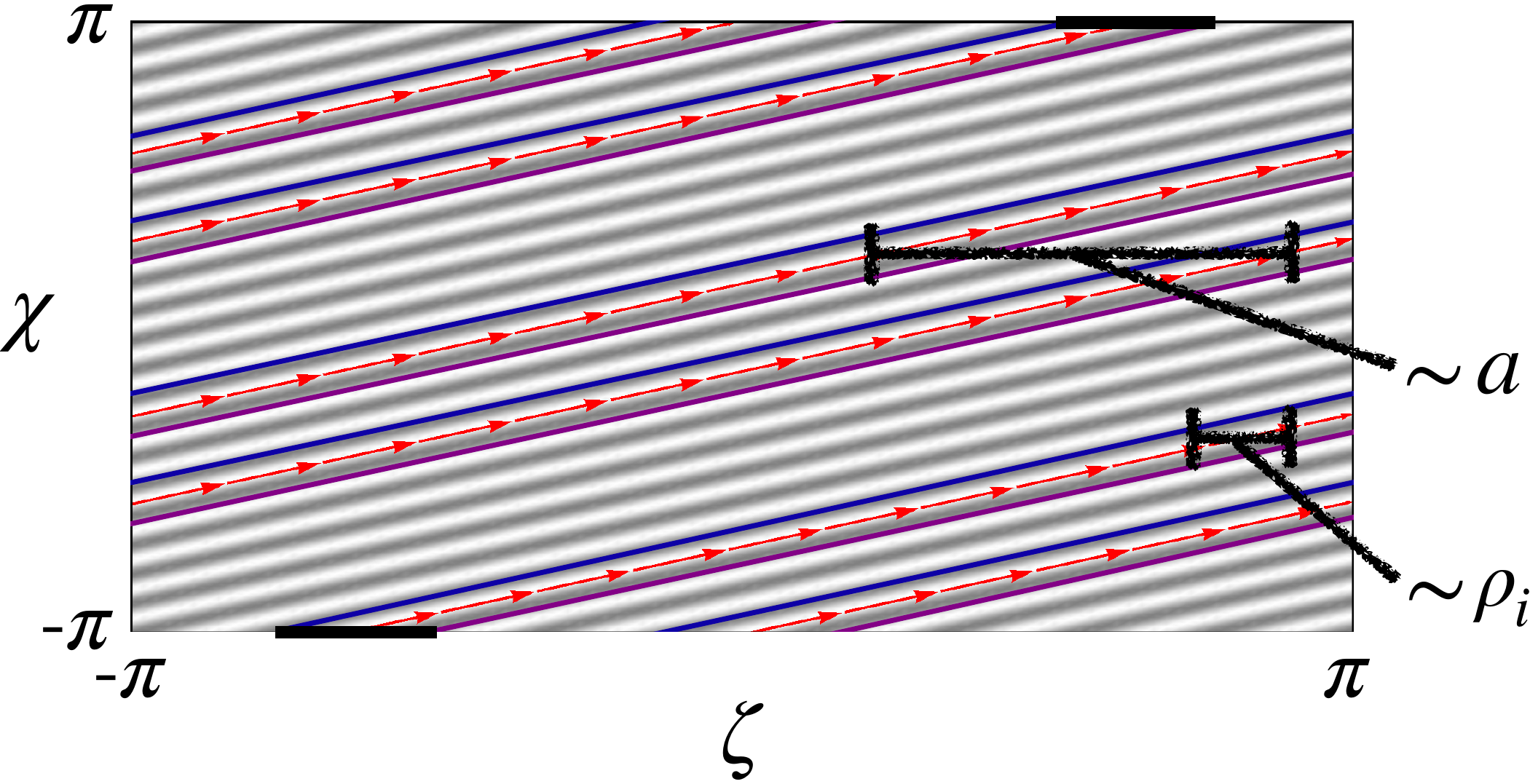}
	\caption{An example flux surface showing two poloidal turns of a magnetic field line (red arrows) and a global perturbation (gray and white stripes) with toroidal and poloidal mode numbers of 9 and 21 respectively. Also shown is a $N_{\text{pol}} = 2$ flux tube (thin blue and purple lines) with the parallel boundary condition applied between the two ends (thick black horizontal lines). As shown, $\chi$ is taken to be a straight-field line poloidal angle.}
	\label{fig:globalConsistency}
\end{figure}

\begin{figure}
	\centering
	\includegraphics[width=0.8\textwidth]{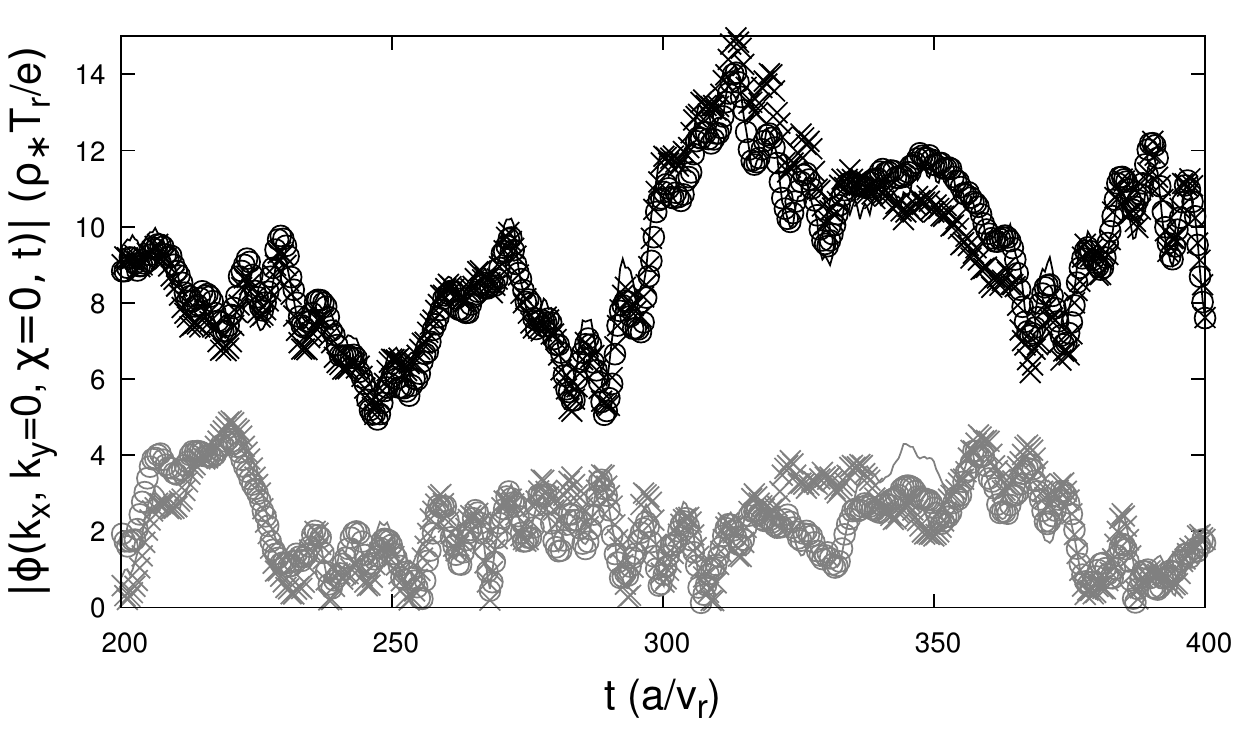}
	\caption{The zonal amplitude of the turbulent electrostatic potential $\phi$ as a function of the time $t$, as calculated by a gyrokinetic simulation with the parameters of tables \ref{tab:resolutions} and \ref{tab:CBC} using kinetic electrons, a binormal domain width $L_{y} = 125 \rho_{i}$, and $N_{\text{pol}} = 3$. The color (i.e. black or gray) indicates two different values of the radial wavenumber (i.e. $k_{x} \rho_{i} \in \{ 0.05, 0.1 \}$ respectively) and the symbol (i.e. line, circle, or cross) indicates the three different toroidal locations at the outboard midplane (i.e. $\chi \in \{- 2 \pi, 0, 2 \pi \}$). Here the reference velocity $v_{r} \equiv \sqrt{T_{r}/m_{r}}$, temperature $T_{r}$, and mass $m_{r}$ are taken to be the ion values and $e$ is the proton charge.}
	\label{fig:zonalFlows}
\end{figure}

\begin{table}
	\centering
	\begin{tabular}{c||c|c||c|c }
		Coordinate    & \multicolumn{2}{c||}{Grid range}
		& \multicolumn{2}{c}{Number of gridpoints}                \\
		\hline
		& Adiabatic & Kinetic & Adiabatic & Kinetic  \\
		\hline
		$x / \rho_{r}$ & $[ - 100, 100 )$ & $[ - 65, 65 )$ & $256 N_{\text{pol}}$ & $128 N_{\text{pol}}$ \\
		\hline
		$y / \rho_{r}$ & $[ - L_{y}/2, L_{y}/2 ) / \rho_{r}$ & $[ - L_{y}/2, L_{y}/2 ) / \rho_{r}$ & $64 L_{y} / L_{y 0}$ & $32 L_{y} / L_{y 0}$ \\
		\hline
		$\chi$ & $[ - N_{\text{pol}} \pi, N_{\text{pol}} \pi )$ & $[ - N_{\text{pol}} \pi, N_{\text{pol}} \pi )$ & $18 N_{\text{pol}}$ & $18 N_{\text{pol}}$ \\
		\hline
		$v_{||} / v_{s}$ & $[ - 3, 3 ]$ & $[ - 3, 3 ]$ & $32$ & $32$ \\
		\hline
		$\sqrt{\mu / (T_{s} / B_{r})}$ & $( 0, 3 )$ & $( 0, 3 )$ & $8$ & $8$ \\
		\hline
		$t/(a/v_{r})$ & $[500, 1500]$ & $[150, 800]$ & \multicolumn{2}{c}{timestep $<$ CFL limit} \\
	\end{tabular}
	\caption{The nominal GENE \citep{JenkoGENE2000} coordinate grids used for the scans in $L_{y}$ and $N_{\text{pol}}$, where $L_{y 0} \equiv 125 \rho_{r}$ is a common value used by the community. Note that all grids are equally spaced, the reference length $a$ is the tokamak minor radius, the reference magnetic field $B_{r}$ is the toroidal field in the flux tube at $R_{0}$, $R_{0}$ is the average major radius of the flux tube, and the CFL condition is explained in \citet{CourantCFL1967}. The other reference quantities indicated by the subscript $r$ are taken to be the ion quantities (except for the electron-scale simulation in figure \ref{fig:shearLayers}) and $v_{s} = \sqrt{2 T_{s} / m_{s}}$ is the thermal velocity of species $s$.}
	\label{tab:resolutions}
\end{table}

\begin{table}
	\centering
	\begin{tabular}{c|c||c|c}
		Parameter & Value & Parameter & Value \\
		\hline
		Minor radius of flux tube, $x_{0}/a$ & 0.54 & Major radius, $R_{0}/a$ & $3.0$ \\
		\hline
		Safety factor, $q_{0}$ & $1.4$ & Magnetic shear, $\hat{s}$ & $0.8$ \\
		\hline
		Temperature gradient, $a/L_{Ts}$ & $2.3$ & Density gradient, $a/L_{ns}$ & $0.733$ \\
		\hline
		Ion-e\textsuperscript{-} mass ratio, $m_{i}/m_{e}$ & $3672$ & Ion-e\textsuperscript{-} temperature ratio, $T_{i}/T_{e}$ & $1.0$ \\
		\hline
		Effective ion charge, $Z_{eff}$ & $1.0$ & Collision frequency, $\nu_{s s'}$ & $0.0$ \\
		\hline
		4\textsuperscript{th} order $v_{||}$ hyperdiffusion, $\epsilon_{v||}$ & $0.2$ & 4\textsuperscript{th} order $z$ hyperdiffusion, $\epsilon_{c}$ & $0.05$ \\
	\end{tabular}
	\caption{The nominal Cyclone Base Case \citep{DimitsCBC2000} parameters used for the scans in $L_{y}$ and $N_{\text{pol}}$. Simulations with kinetic electrons use $\epsilon_{v||} = \epsilon_{c} = 0.5$ \citep{PueschelHyperDiff2010} and a plasma beta of $\beta = 0.001$ for numerical reasons.}
	\label{tab:CBC}
\end{table}

While $N_{\text{pol}}$ is free to be any positive integer, it has become overwhelmingly standard in the fusion community to use $N_{\text{pol}} = 1$ (e.g. figure \ref{fig:fluxTubeGeo}). While this is likely due to the desire to minimize the computational cost of simulations, there also have been concerns about the validity of flux tubes with $N_{\text{pol}} > 1$. For example, such simulations are often not ``globally consistent'' \citep{ScottFluxTubeConsistency1998}. This means that, when $N_{\text{pol}} > 1$, the narrow flux tube can permit Fourier modes that do not exist on the full, doubly-periodic flux surface. 
However, in the true $\rho_{\ast} \rightarrow 0$ limit for which gyrokinetics is valid, the charged particles have no way of knowing that they are on a doubly-periodic flux surface. Particles can never communicate information between the different poloidal turns by moving in the toroidal direction because the distance separating the different poloidal turns is proportional to the minor radius $a$, while the turbulent correlation length is proportional to the ion gyroradius $\rho_{i}$ (see figure \ref{fig:globalConsistency}). Thus, the distance separating the poloidal turns is an asymptotically large number of turbulent correlation lengths, so the particles will perceive the perpendicular direction within the flux surface as infinite in the $\rho_{\ast} \equiv \rho_{i}/a \rightarrow 0$ limit. This means that it is not relevant which modes are allowed on a doubly-periodic flux. While we should still be concerned about respecting any periodicity in the parallel direction (i.e. do not use a domain with $N_{\text{pol}} = 2$  to model a $q = 1$ surface), violating global perpendicular periodicity will only introduce an error that is small in $\rho_{\ast} \ll 1$. This is intuitive --- {\it global} consistency should be unimportant in {\it local} simulations. Another concern is that flux tubes with $N_{\text{pol}} > 1$ include multiple locations at the same poloidal angle, which must all have the same value of the zonal flows (i.e. fluctuations with a binormal wavenumber $k_{y} = 0$). Fortunately, it appears that the physics of the gyrokinetic model ensures that this is fairly well satisfied in the simulations we have run (e.g. figure \ref{fig:zonalFlows}). However, we should point out that we have no proof that this must {\it always} be the case, so it remains an open issue.

\begin{figure}
	\centering
	\begin{minipage}{\textwidth}
		\centering
		$\vcenter{\hbox{\includegraphics[width=0.2\textwidth]{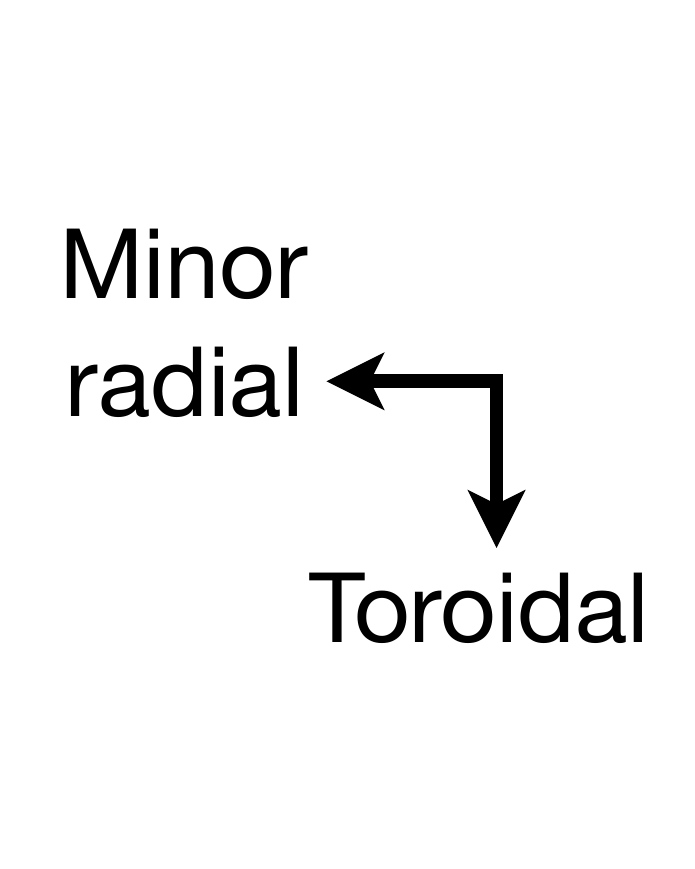}}}$
		\hspace*{-0.1in}
		$\vcenter{\hbox{\includegraphics[width=0.2\textwidth]{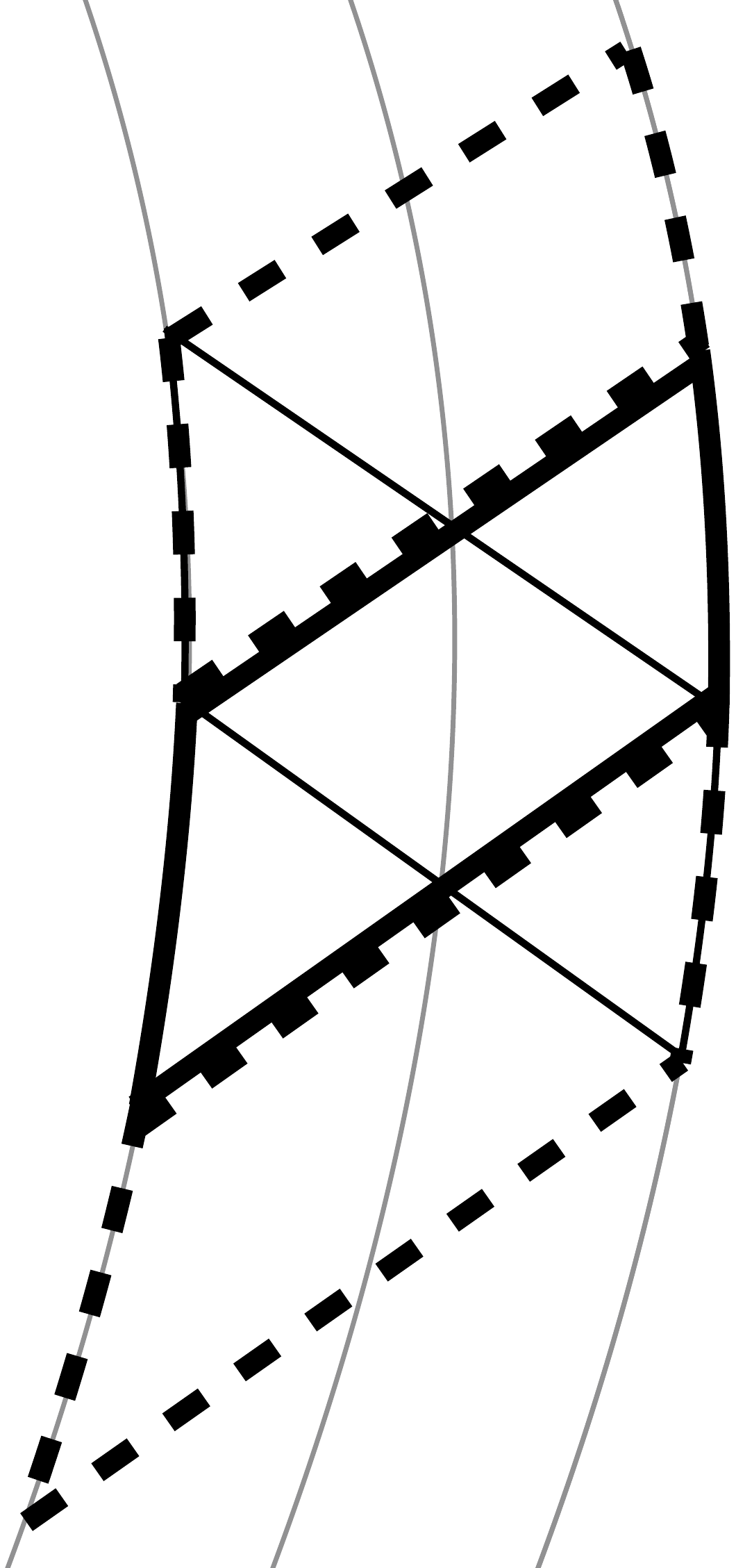}}}$
	\end{minipage}
	\caption{The shape of the parallel boundary (thin solid black parallelogram) on the inboard midplane of the example flux tube shown in figure \ref{fig:fluxTubeGeo}. The parallel boundary condition is supplied by shifting the other end of the flux tube (thick solid black parallelogram) toroidally until it is centered on the first and then making periodic copies offset in the toroidal direction (two thick dashed black parallelograms). The flux surfaces are also shown (thin solid gray lines).}
	\label{fig:boundaryGeo}
\end{figure}

Unlike with the perpendicular boundary conditions, substituting statistical periodicity with exact periodicity across the parallel boundary is made more complicated by the effect of magnetic shear. In local simulations, the safety factor profile is approximated to be linear across the flux tube, i.e. 
\begin{align}
q(x) = q_{0} + \frac{d q}{d x} ( x - x_{0} ) = q_{0} \left( 1 + \hat{s} \frac{x-x_{0}}{x_{0}} \right) , \label{eq:safetyFactor}
\end{align}
where $q$ is the safety factor, $x \equiv r$ is the minor radial coordinate within the domain, $\hat{s} \equiv (x_{0} / q_{0}) d q / d x$ is the magnetic shear, and the subscript $0$ indicates the quantity is evaluated at the center of the radial domain. This means that the two ends of the flux tube are deformed into parallelograms that have opposite tilts. Hence, the shapes do not overlap nicely, which must be resolved by the parallel boundary condition (see figure \ref{fig:boundaryGeo}). In essence, the parallel boundary condition determines how the magnetic field lines are connected between these two parallelograms. The twist-and-shift condition typically implemented by gyrokinetic codes starts by using axisymmetry to justify toroidally shifting the two parallelograms until they are centered on one another \citep{BeerBallooingCoordinates1995}. Then, again using axisymmetry, we can make periodic copies of the flux tube, shifted in the toroidal direction, and completely cover the opposing parallelogram (see figure \ref{fig:boundaryGeo}). Thus, any turbulent structure that extends past one end of the domain will be copied into the other end while maintaining the twisting caused by magnetic shear\footnote{For a mathematical explanation of the parallel boundary condition in real space see appendix \ref{app:aspectRatio} or for a Fourier-space formulation see \citet{BeerBallooingCoordinates1995}.}.

\begin{figure}
	\centering
	\begin{minipage}{\textwidth}
		\centering
		$\vcenter{\hbox{\includegraphics[width=0.4\textwidth]{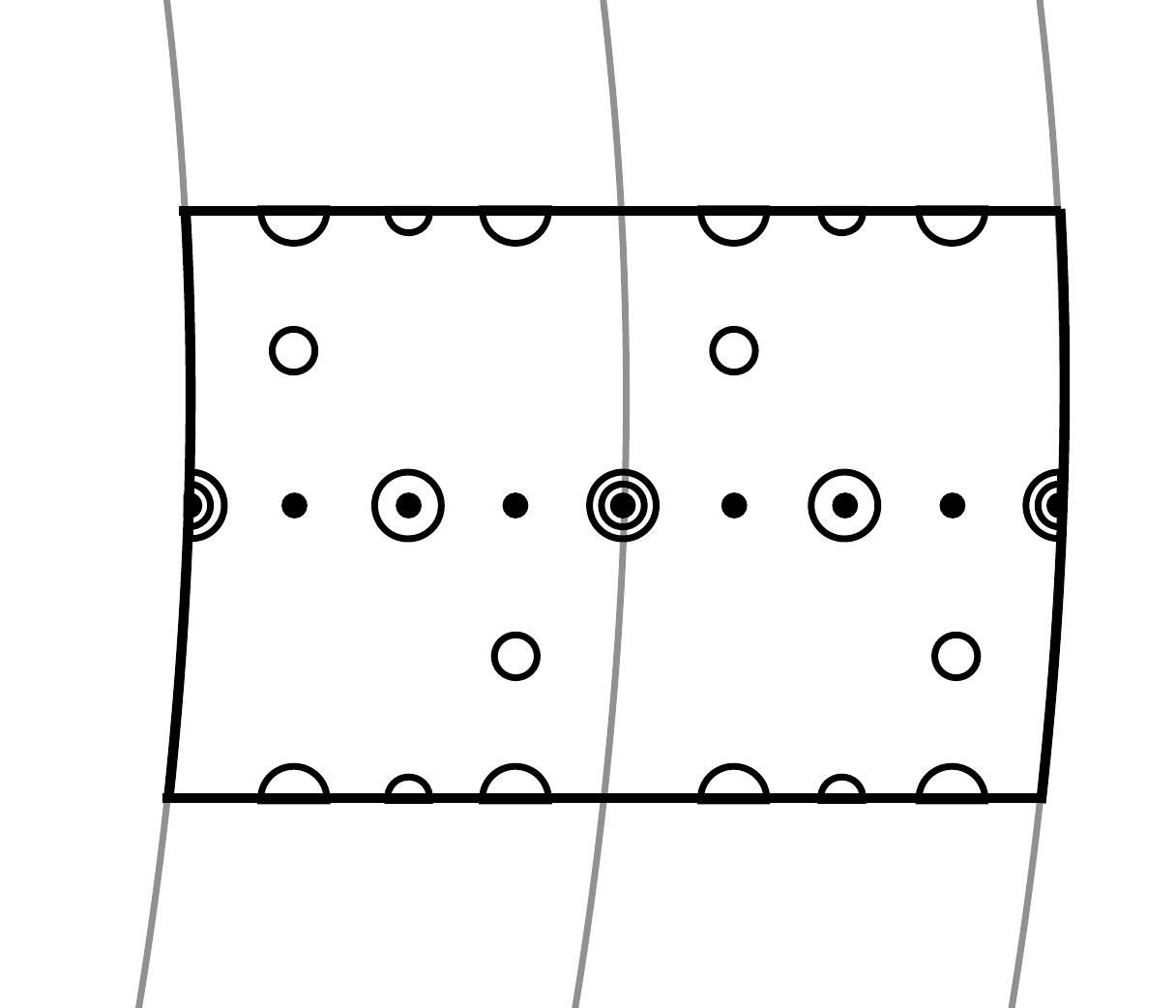}}}$
		\hspace*{-0.2in}
		$\vcenter{\hbox{\includegraphics[width=0.2\textwidth]{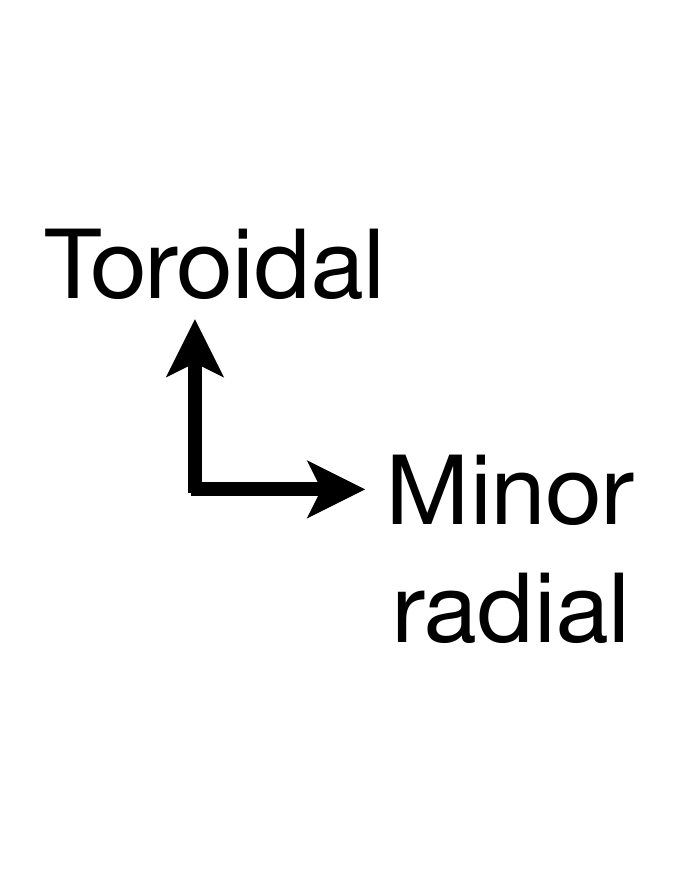}}}$
	\end{minipage}
	\caption{A Poincar\'{e} puncture plot at the outboard midplane for the flux tube from figures \ref{fig:fluxTubeGeo} and \ref{fig:boundaryGeo}, which has $N_{\text{pol}} = 1$. Following a magnetic field line that starts at different radial positions (smallest filled circles) will take you to different toroidal locations after one poloidal circuit of the device (smaller empty circles) and two poloidal circuits (large empty circles). Two radial locations have field lines that close on themselves after passing through the parallel boundary just once, so $N_{\text{si}} = 2$.}
	\label{fig:puncturePlot}
\end{figure}

One consequence of this boundary condition is that it forces at least one flux surface in the domain to have field lines that close on themselves after going through the parallel boundary just once or, equivalently, after just $N_{\text{pol}}$ poloidal turns. Given that we have chosen to center the parallelograms with respect to one another, the flux surface in the center of the radial domain will always be such a surface (see figure \ref{fig:puncturePlot}). However, figures \ref{fig:boundaryGeo} and \ref{fig:puncturePlot} show that there can be more than one --- sometimes there are many. We will indicate the total number of surfaces with field lines that close on themselves after just one pass through the parallel boundary with $N_{\text{si}}$. The subscript $si$ stands for ``self-interaction'' because these surfaces will exhibit the strongest parallel self-interaction. Now if $N_{\text{pol}} = 1$, the parallel boundary condition will be applied after one poloidal turn, so these field lines will close after just one poloidal turn. Accordingly, we will call them ``pseudo-integer'' surfaces --- they close on themselves after one poloidal turn like a normal integer surface, but they are usually an artifact of the boundary condition and don't actually correspond to integer flux surfaces in the real device. Additionally, we see that there will be ``pseudo-rational'' surfaces at other radial locations, where the field lines exactly close on themselves after two or more poloidal turns. For example, the ``pseudo-half-integer'' surfaces close on themselves after two poloidal turns and will always occur midway between two neighboring pseudo-integer surfaces. Alternatively, if $N_{\text{pol}}$ is increased from $1$ to $2$, all of the pseudo-integer surfaces will be replaced by pseudo-half-integer surfaces. Thus, the lowest order rational surfaces in an $N_{\text{pol}} = 2$ simulation will be pseudo-half-integer and there will be $N_{\text{si}}$ of them.

These pseudo-rational surfaces are also involved in an important constraint that arises from the combination of the parallel and radial boundary conditions --- the aspect ratio of the domain becomes discretized (i.e. the radial domain width $L_{x}$ divided by the binormal domain width $L_{y}$). This can be found from a derivation in real space (see appendix \ref{app:aspectRatio}), a derivation in Fourier-space \citep{BeerBallooingCoordinates1995}, or by the following explanation. First note that radial periodicity implies an equivalence between pairs of field lines, one on each of the two radial boundaries. In other words, any field line at $x = x_{0} - L_{x}/2$ has a matching field line at $x = x_{0} + L_{x}/2$ with which it is equivalent. These field lines are the same and must remain matched as you move in the parallel direction. Otherwise, a particle moving purely along a field line would find itself transported {\it across} field lines. However, from figures \ref{fig:boundaryGeo} and \ref{fig:puncturePlot}, we see that the twist-and-shift condition, which is applied at the parallel ends of the domain, connects different field lines at different toroidal locations. This can cause problems. If we are not careful, the twist-and-shift boundary condition will connect the two field lines from a matching pair to two other field lines that are not matching. This would cause spurious cross-field transport. We must ensure that, given any matching pair of field lines, twist-and-shift always connects them to two other field lines that also form a matching pair. From figures \ref{fig:boundaryGeo} and \ref{fig:puncturePlot}, we see that this will only be achieved if
\begin{align}
\Delta \zeta = N_{\text{si}} L_{\zeta} , \label{eq:zetaShift}
\end{align}
where $\Delta \zeta$ is the total shift in toroidal angle caused by magnetic shear along the length of the flux tube, $N_{\text{si}}$ is some integer, and $L_{\zeta}$ is the width of the flux tube in toroidal angle such that $L_{\zeta} = 2 \pi$ corresponds to the full toroidal domain. For the example shown in figures \ref{fig:boundaryGeo}, we can see that $N_{\text{si}} = 2$ because the radial boundaries of the thick solid black parallelogram are offset in the toroidal direction by $1 \times L_{y}$ and this represents half the total shift (i.e. the shift in the range $\chi \in [ - \pi N_{\text{pol}}, 0 ]$, but not $\chi \in [ 0, \pi N_{\text{pol}} ]$). Additionally, figure \ref{fig:puncturePlot} shows that, for this example, the twist-and-shift condition connects every field line on the radial boundary to itself. This trivially ensures that matching pairs remain matching as particles move across the parallel boundary. If $N_{\text{si}} = 1$, the field lines on the radial boundary would no longer close on themselves, but instead to a different matching pair offset in the $y$ direction by $L_{y}/2$.  

To see how equation \refEq{eq:zetaShift} discretizes the domain aspect ratio, we invert the definition of the binormal coordinate
\begin{align}
y(x,\zeta,\chi) \equiv \frac{x_{0}}{q_{0}} ( q(x) \chi - \zeta ) \label{eq:yDef}
\end{align}
(where $\chi$ is a straight-field line poloidal angle) and use equation \refEq{eq:safetyFactor} to calculate the toroidal shift across the sheared domain to be
\begin{align}
\Delta \zeta &\equiv \zeta (x = x_{0} + \frac{L_{x}}{2}, y = y_{0}, \chi = 2 \pi N_{\text{pol}}) - \zeta (x = x_{0} - \frac{L_{x}}{2}, y = y_{0}, \chi = 2 \pi N_{\text{pol}}) \\
&= 2 \pi N_{\text{pol}} \frac{d q}{d x} L_{x} = 2 \pi N_{\text{pol}} \frac{q_{0}}{x_{0}} \hat{s} L_{x} , \label{eq:toroidalShift}
\end{align}
where $y_{0}$ is any arbitrary binormal location. Since
\begin{align}
L_{y} = \frac{x_{0}}{q_{0}} L_{\zeta} \label{eq:binormalWidths}
\end{align}
follows from evaluating equation \refEq{eq:yDef} at constant $x$ and $\chi$, we can use equations \refEq{eq:zetaShift} and \refEq{eq:toroidalShift} to see that the possible values for the radial domain width are
\begin{align}
L_{x} = \frac{N_{\text{si}}}{2 \pi N_{\text{pol}} | \hat{s} |} L_{y} . \label{eq:aspectRatio}
\end{align}
Thus, we find that $L_{x}$ must be a integer multiple of the distance between lowest-order pseudo-rational surfaces. Importantly, we see that if the magnetic shear is very small, the radial width will be forced to be large because you cannot decrease $N_{\text{si}}$ below $1$. This causes simulations with very low, but finite magnetic shear to become expensive. Equation \refEq{eq:aspectRatio} also shows that $N_{\text{si}}$ can be used to specify the radial width of the domain. In fact, $N_{\text{si}}$ is a common input parameter to gyrokinetic codes and goes by various names such as \texttt{nexc} in GENE, \texttt{jtwist} in GS2, \texttt{IKXSPACE} in GKW, and \texttt{m\_j} in GKV.

Lastly, it is important to note that the pseudo-rational surfaces created by the parallel boundary condition {\it can} be physical \citep{WaltzShearLayers2006, Weikl2018, AjaySelfInteraction2020}. For example, a $q=3$ surface really closes on itself after one poloidal turn and should be modeled using $N_{\text{pol}} = 1$, while a $q=5/2$ surface should use $N_{\text{pol}} = 2$. Moreover, the physics occurring at these surfaces may be crucial for modeling certain behavior such as microtearing modes \citep{HazeltineMicrotearing1975, GuttenfelderMicrotearing2011}. However, to accurately model the overall impact of self-interaction from rational surfaces, the spacing between pseudo-rational surfaces must also be correct. For example, the domain shown in figure \ref{fig:puncturePlot} has $N_{\text{si}} = 2$ pseudo-integer surfaces, so, to be physical, the radial width of the simulation domain must correspond to twice the distance between the actual integer surfaces in the device\footnote{Or more precisely, twice the distance between the integer surfaces in the device {\it if the safety factor profile was given by equation \refEq{eq:safetyFactor}}.}. To put it another way, the simulation must have the same number of ion gyroradii separating the integer surfaces as the experiment does. Unfortunately, accomplishing this turns out to require a simulation domain that corresponds to the full flux surface (as is discussed further at the end of section \ref{sec:resStudy}). To see this, take the change in safety factor across the simulation domain $\Delta q = L_{x} d q / d x$ and set it equal to the number of integer surfaces in the simulation $\Delta q = N_{\text{si}}$. Then, substituting equations \refEq{eq:binormalWidths} and \refEq{eq:aspectRatio} with $N_{\text{pol}} = 1$ shows that $L_{\zeta} = 2 \pi$ (i.e. the toroidal domain size is equal to the entire toroidal domain). Thus, we see that accurately modeling self-interaction on rational surfaces requires full flux surface simulations with radial domain widths on the scale of the tokamak minor radius $a$. Flux tubes of this size offer little computational savings compared to global simulations, but if we try to get by with a smaller domain of $L_{\zeta} = 2 \pi/N$, it will have a radial density of pseudo-integer surfaces that is $N$ times larger than reality.

\section{Neglecting pseudo-rational surfaces}
\label{sec:rationalSurf}

While the pseudo-rational surfaces created by the parallel boundary condition are unphysical unless the domain size is very large, this isn't always a problem. Typically gyrokinetic simulations are used to model plasma far from low-order rational surfaces. To do this accurately we do {\it not} need to ensure that the simulation has the correct spacing of rational surfaces, we just need to ensure that their presence is negligible. Broadly-speaking, this can be accomplished by verifying that the results of the simulation do not depend on the spacing of pseudo-rational surfaces. From equation \refEq{eq:aspectRatio}, we see that the spacing is given by
\begin{align}
\frac{L_{x}}{N_{\text{si}}} = \frac{L_{y}}{2 \pi N_{\text{pol}} | \hat{s} |} , \label{eq:surfDensity}
\end{align}
so we can vary either $L_{y}$ or $N_{\text{pol}}$ and test for convergence. Additionally, we can specifically check the low-order pseudo-rational surfaces to see if they are exhibiting any unusual behavior. The low-order pseudo-rational surfaces are only distinct from the other surfaces in that they close on themselves after a small number of poloidal turns. If the parallel correlation length of the turbulence is sufficiently long, an eddy can interact with itself along the field line, thereby altering its statistical properties. Hence, to ensure that their presence is not affecting things, we can verify that the flux tube has statistically homogeneous turbulence in the perpendicular plane.

\begin{figure}
	\centering
	\includegraphics[width=\textwidth]{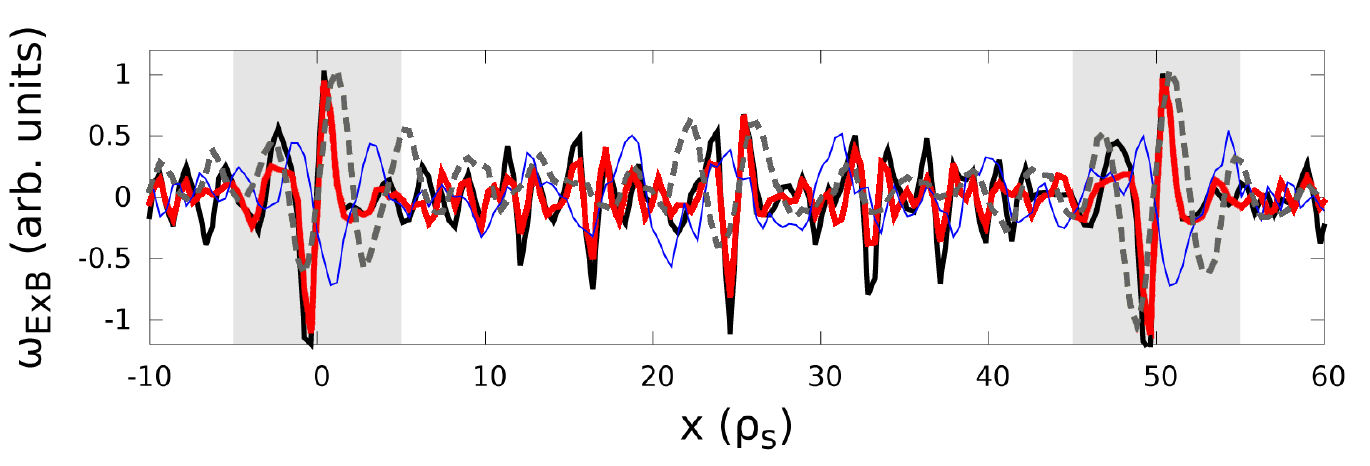}
	\caption{The time-averaged zonal $E \times B$ shearing rate (i.e. $\partial^{2} \phi / \partial x^{2}$ averaged over time and $y$) as a function of minor radial location within the flux tube for slab simulations (though toroidal simulations with kinetic electrons look similar \citep{Dominski15, AjaySelfInteraction2020}). Data are shown for ion scale simulations using kinetic electrons with $m_i/m_e = 3672$ (black thick solid), kinetic electrons with $m_i/m_e = 918$ (red thick solid), and adiabatic electrons (gray thick dashed). An electron scale simulation is shown with adiabatic ions (blue thin solid). The light gray regions indicate the portion significantly affected by pseudo-integer surfaces, while pseudo-half-integer surfaces are located halfway in-between. The parameters used are given in tables \ref{tab:resolutions} and \ref{tab:CBC}, except $a/L_{Ts} = 10$, $a/L_{ns} = 2.0$, and several of the geometry parameters are not needed. Note the reference magnetic field $B_{r}$ and reference length $a$ become arbitrary in slab geometry.}
	\label{fig:shearLayers}
\end{figure}

For example, the observation of localized, steady $E \times B$ flow shear layers \citep{CandyShearLayers2005, WaltzShearLayers2006, Dominski15, Weikl2018, AjaySelfInteraction2020} is a clear indicator that pseudo-rational surfaces may be adversely affecting the accuracy of many present-day simulations. These structures are a general consequence of self-interaction and are a fairly universal in conventional tokamak simulations with kinetic electrons. Although it is difficult to see them in tokamak simulations with adiabatic electrons, figure \ref{fig:shearLayers} shows that they can be found by using a slab geometry (instead of toroidal geometry). 
While they have been identified as an electron-scale phenomenon {\it in linear simulations} \citep{Dominski15}, in nonlinear simulations they manifest as fairly narrow ion-scale structures. Figure \ref{fig:shearLayers} demonstrates that their width does not scale with the electron mass in slab ion-scale simulations and the same behavior has been seen in previous toroidal ion-scale simulations \citep{Dominski17}. Furthermore, figure \ref{fig:shearLayers} shows that similar structures can be found at electron-scales in slab Electron Temperature Gradient (ETG) simulations.

These $E \times B$ flow shear layers are particularly interesting because they are constant in time --- i.e. the plasma is spontaneously moving momentum around within the flux tube in order to modify the steady background flow profile. Like {\it large-scale} steady $E \times B$ flows, the tokamak will set up a parallel flow in order to convert the $E \times B$ flow layers into a purely toroidal flow \citep{AbelGyrokineticsDeriv2012}. Such flow represents intrinsic rotation, which is generally prohibited by an underlying symmetry of the gyrokinetic model \citep{PeetersMomTransSym2005, ParraUpDownSym2011, SugamaUpDownSym2011}. Thus, there must be a symmetry-breaking mechanism present. One plausible mechanism is variation in turbulence characteristics \citep{ParraIntrinsicRotTheory2015}. This could be introduced by the existence of pseudo-rational surfaces (e.g. the turbulence on integer surfaces is different than on the neighboring irrational surfaces). Normally, this symmetry-breaking effect is small in $\rho_{\ast} \ll 1$ because the gradual change in background gradients is the usual cause for the variation in turbulence characteristics. However, the variation caused by the pseudo-integer surfaces occurs on the scale of the gyroradius, so it is {\it not} small in $\rho_{\ast}$. There are only two physical effects that cause variation in turbulence characteristics to break the symmetry (see Section 9 of \citet{ParraIntrinsicRotTheory2015}): finite gyroradius effects and the radial magnetic drifts. Since figure \ref{fig:shearLayers} shows the shear layers persist in slab geometry, we believe that the finite gyroradius effect is the dominant symmetry breaking effect. This is intuitive as the symmetry breaking occurs on a small scale and appears to be dipolar (i.e. a region of positive momentum flux is always next to a region of negative), so the radial drift orbits should average over it more effectively because they are significantly larger than the gyroradius.

\begin{figure}
	\centering
	\includegraphics[width=\textwidth]{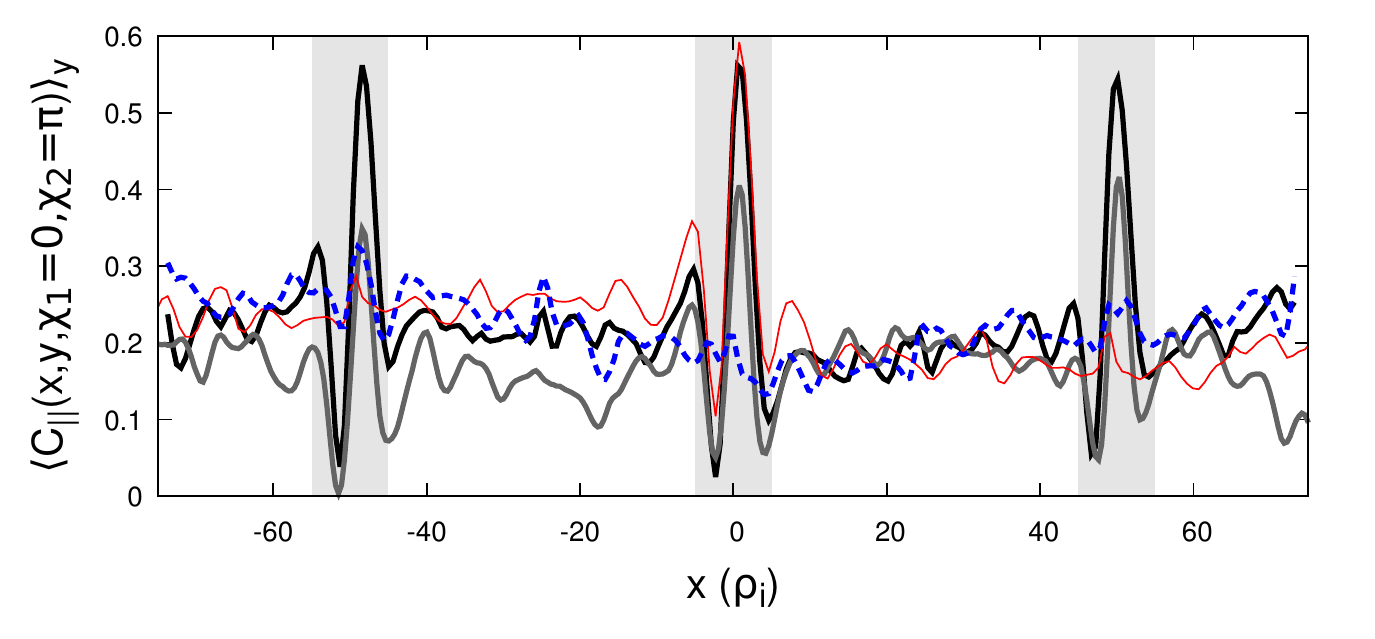}
	\caption{The parallel correlation between inboard and outboard locations following a field line at a given minor radial location within the flux tube. The base simulation has kinetic electrons with $N_{\text{pol}} = 1$ and $L_{y} = 250 \rho_{i}$ (black thick solid). The other three curves are produced by changing the base simulation to either have adiabatic electrons (gray thick solid), $L_{y} = 500 \rho_{i}$ (red thin solid), or $N_{\text{pol}} = 2$ (blue thick dashed). Note the $L_{y} = 500 \rho_{i}$ simulation has only a single pseudo-integer surface, which occurs at $x = 0$. The $N_{\text{pol}} = 2$ simulation has no pseudo-integer surfaces, but has pseudo-half-integer surfaces at $x = \{ 0, \pm 25, \pm 50, \pm 75 \} \rho_{i}$}
	\label{fig:parCorr}
\end{figure}

Importantly, these flow shear layers are just one possible symptom of self-interaction at pseudo-rational surfaces. More generally, we can investigate spurious correlations by looking at the two-point parallel correlation function
\begin{align}
C_{||} \left( x, y, \chi_{1}, \chi_{2} \right) \equiv \frac{\left\langle \phi_{NZ} \left( x, y, \chi_{1}, t \right) \phi_{NZ} \left( x, y, \chi_{2}, t \right) \right\rangle_{t}}{\sqrt{ \left\langle \phi_{NZ}^{2} \left( x, y, \chi_{1}, t \right) \right\rangle_{t} \left\langle \phi_{NZ}^{2} \left( x, y, \chi_{2}, t \right) \right\rangle_{t} }} ,
\end{align}
where the subscript $NZ$ signifies the non-zonal portion of the quantity and $\left\langle \ldots \right\rangle_{u}$ indicates an average over any coordinate $u$. The quantity $C_{||}$ indicates the degree of correlation between two points $\chi_{1}$ and $\chi_{2}$ on the same field line (i.e. at constant $x$ and $y$). Figure \ref{fig:parCorr} shows the $y$-averaged correlation between adjacent outboard and inboard midplanes $\left\langle C_{||} ( x, y, \chi_{1} = 0, \chi_{2} = \pi ) \right\rangle_{y}$ for various toroidal simulations. Note that all simulations with $N_{\text{pol}} = 1$ have sharp spikes in the parallel correlation at the locations of pseudo-integer surfaces. Thus, even though toroidal simulations with adiabatic electrons did not display flow shear layers at pseudo-integer surfaces, they {\it do} exhibit spikes in parallel correlation. We see that increasing $L_{y}$ increases the spacing between pseudo-integer surfaces, in accordance with equation \refEq{eq:surfDensity}. However, we also notice that neither the width, nor the height of the spikes changes with $L_{y}$. Thus, as we increase $L_{y}$ the regions affected by pseudo-integer surfaces occupy a smaller and smaller fraction of the domain. This indicates one way to eliminate the effect of pseudo-rational surfaces--- increase $L_{y}$ until convergence is achieved. If $L_{y}$ is sufficiently large, the pseudo-rational surfaces will have a negligible impact on all volume-averaged quantities. Effectively we are ``diluting'' away their influence.

Figure \ref{fig:parCorr} also shows a second strategy to eliminate pseudo-integer surfaces --- increase $N_{\text{pol}}$ until convergence is achieved. If the flux tube is longer, a turbulent eddy has to remain correlated over a longer distance in order to ``bite its own tail.'' In other words, the $N_{\text{pol}} = 2$ has no pseudo-integer surfaces, only pseudo-half-integer surfaces. Accordingly, we see that, compared to the $N_{\text{pol}} = 1$ simulation, the $N_{\text{pol}} = 2$ case has much smaller peaks at its lowest order pseudo-rational surfaces, which are more closely spaced. This suggests that, given the parameters used for these simulations, a small number of poloidal turns should be sufficient to eliminate the effect of pseudo-rational surfaces and achieve convergence. This is consistent with the original convergence study of \citet{BeerBallooingCoordinates1995}, which found that two or three poloidal turns was sufficient to achieve convergence in gyrofluid simulations with adiabatic electrons. However, it was {\it not} obvious that kinetic and adiabatic electron simulations would behave similarly because, in linear simulations, kinetic electrons enable modes that are very extended along the magnetic field line (i.e. ``giant tails'' in the ballooning envelope) \citep{HallatschekGiantTails2005}.

\section{Resolution study}
\label{sec:resStudy}

\begin{figure}
	\hspace{3em} (a)
	
	\begin{center}
		\includegraphics[width=0.8\textwidth]{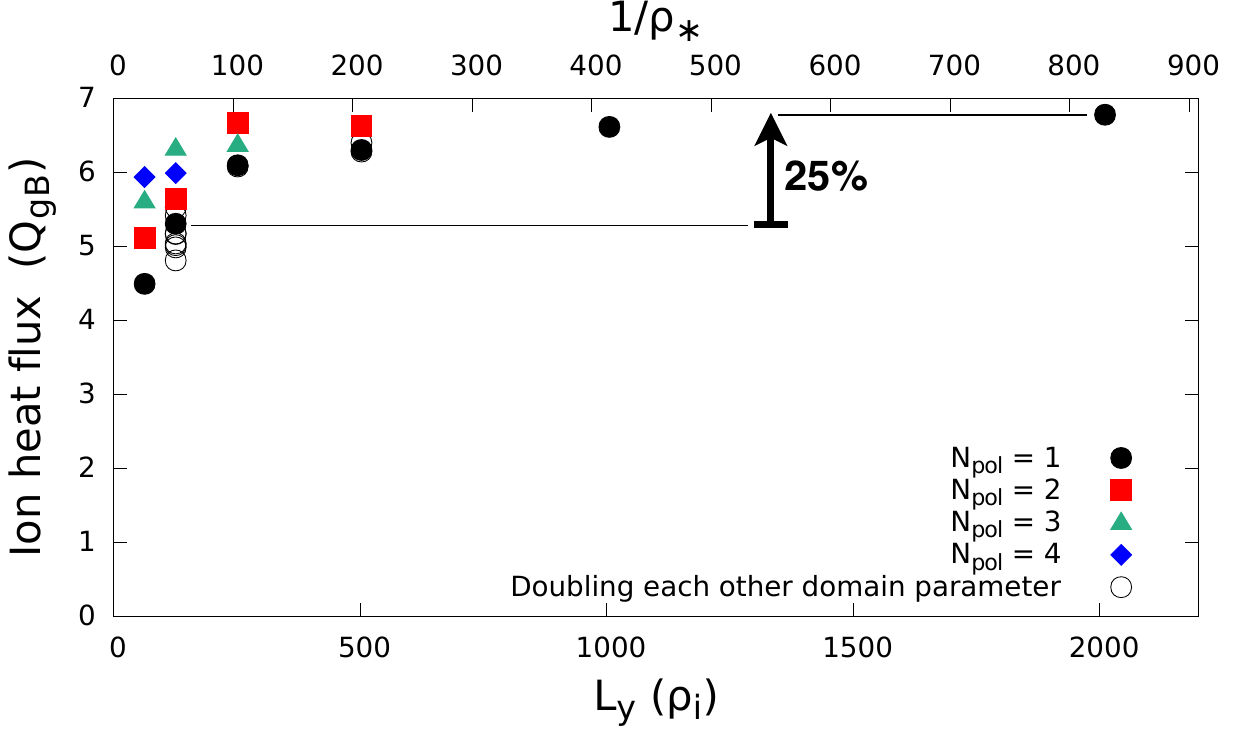}
	\end{center}
	
	\hspace{3em} (b)
	
	\begin{center}
		\includegraphics[width=0.8\textwidth]{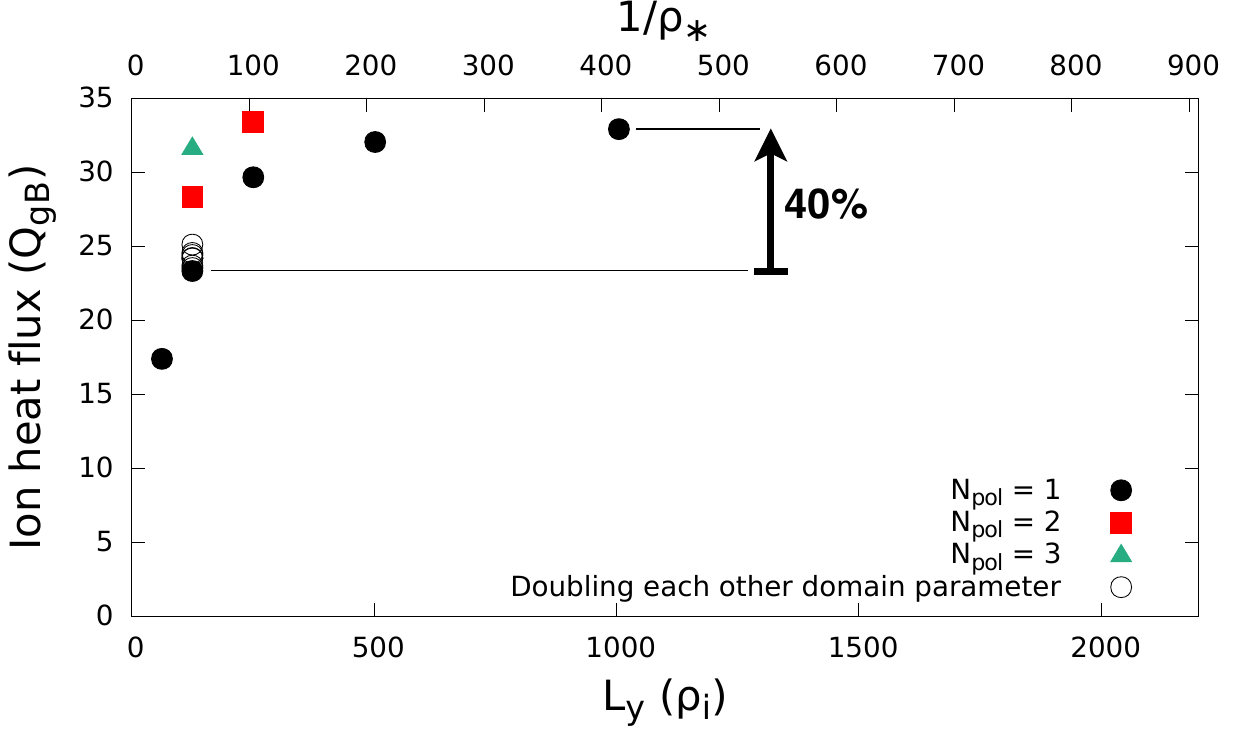}
	\end{center}
	\caption{Resolution studies in $L_{y}$ and $N_{\text{pol}}$ with the parameters of tables \ref{tab:resolutions} and \ref{tab:CBC} using adiabatic electrons (a) and kinetic electrons (b). The open black circles indicate simulations with $N_{\text{pol}} = 1$ that individually double the grid resolution or the domain width of the three other coordinates (i.e. $x$, $v_{||}$ and $\mu$) as well as the grid resolutions in $y$ and $\chi$. The simulation at $L_{y} = 2000 \rho_{i}$ was forced to have double the nominal $x$ domain width (and thus double the number of $x$ gridpoints) due to the aspect ratio quantization of equation \refEq{eq:aspectRatio}. The top axis shows the value of $\rho_{\ast}$ for which a given $L_{y}$ corresponds to the full flux surface.}
	\label{fig:resStudy}
\end{figure}

We will now use the gyrokinetic code GENE \citep{JenkoGENE2000, GoerlerGlobalGENE2011} to perform a resolution study in $L_{y}$ and $N_{\text{pol}}$, the two strategies to eliminate pseudo-integer surfaces. Figures \ref{fig:resStudy}(a) and \ref{fig:resStudy}(b) show the results using standard Cyclone Base Case parameters \citep{DimitsCBC2000} with adiabatic electrons and kinetic electrons respectively (see table \ref{tab:CBC}). The simulations with adiabatic electrons are considerably less expensive, so we are able to perform a more complete parameter scan and more rigorously ensure adequate resolution (see table \ref{tab:resolutions}). Regardless, both display qualitatively similar behavior.

For adiabatic electrons, figure \ref{fig:resStudy}(a) shows that the fully converged value of the ion heat flux appears to be a bit less than $7$ in gyroBohm units ($Q_{gB} = \rho_{\ast}^{2} v_{r} n_{i} T_{r}$, where $n_{i}$ is the ion number density). You can achieve such convergence by simply increasing $L_{y}$ to very large values while maintaining $N_{\text{pol}} = 1$. The same result can also be achieved by increasing $N_{\text{pol}}$, which somewhat alleviates our concerns about the validity of using $N_{\text{pol}} > 1$ (see Section \ref{sec:parallelBC}). However, this is only true when you are already using a sufficiently large value of $L_{y}$. If $L_{y}$ isn't large enough, the simulation won't be fully converged even if $N_{\text{pol}} \rightarrow \infty$. This is for the same reason that we need a sufficiently large domain in the $x$ direction --- self-interaction in the perpendicular direction. Completely independent of parallel self-interaction, you still need to ensure that the turbulent correlation length in the $y$ direction is much smaller than the domain width in the $y$ direction (as explained in section \ref{sec:intro}). If this is not satisfied, convergence will be achieved as $N_{\text{pol}} \rightarrow \infty$ (because the parallel self-interaction will vanish), but it won't converge to the same value as when the binormal self-interaction is also eliminated. Note in figure \ref{fig:resStudy} that the simulations that are not fully resolved in either $L_{y}$ or $N_{\text{pol}}$ have heat fluxes {\it lower} than the converged value. This indicates that both binormal and parallel self-interaction tend to reduce the ability of turbulence to transport energy. This may be because the correlations imposed by self-interaction reduce the degrees of freedom accessible to the turbulence, preventing it from behaving in ways that would more effectively drive transport.

Figure \ref{fig:resStudy}(a) shows that using $L_{y} \approx 125 \rho_{i}$ and $N_{\text{pol}} = 1$, as is common in the community \citep{WatanabeFluxTubeTrain2015}, under-predicts the ion heat flux by about $25\%$ in adiabatic electron simulations. While this is already concerning, figure \ref{fig:resStudy}(b) shows that the effect of self-interaction increases when using kinetic electrons. Instead of a $25\%$ error, we see roughly $40\%$. This is consistent with figure \ref{fig:parCorr}, which showed that the kinetic electron simulation had more pronounced spikes in the parallel correlation function. Moreover, the effect of self-interaction can be even bigger for other physical parameters. For example, \citet{WatanabeFluxTubeTrain2015} shows that taking Cyclone Base Case parameters and lowering the magnetic shear from $\hat{s} = 0.8$ to $0.2$ increases the error in adiabatic electron simulations from $25\%$ to at least $50\%$. Going further, we simulated Cyclone Base Case parameters with adiabatic electrons and $\hat{s} = 0$ and found that changing $N_{\text{pol}} = 1$ to $N_{\text{pol}} = 2$ increased the heat flux by $70\%$. Simulations of stellarators with $\hat{s} = 0$ have even found that turbulence can be stable when using $N_{\text{pol}} = 1$, but unstable when $N_{\text{pol}} = 2$ (see figure 11 of \citet{FaberSelfInteractionStellarator2018}). Clearly, self-interaction through the parallel boundary condition can be a significant effect in local gyrokinetic simulations.

An important practical result is the relative computational cost of the two routes --- increasing $N_{\text{pol}}$ or increasing $L_{y}$. This is briefly considered by \citet{BeerBallooingCoordinates1995} in the context of gyrofluid simulations of a TFTR experimental shot, concluding that ``it appears that faster convergence is obtained by allowing the domain to be longer than a parallel correlation length.'' However, our conclusion from the above Cyclone Base Case simulations is the opposite --- increasing $L_{y}$ while maintaining $N_{\text{pol}} = 1$ was the cheapest way to reach convergence. For example, the $N_{\text{pol}} = 2$, $L_{y} = 250 \rho_{i}$ simulation required roughly double the computational resources of the $N_{\text{pol}} = 1$, $L_{y} = 1000 \rho_{i}$ simulation. This can be understood by studying how the coordinate system grids change in response to increasing $L_{y}$ versus $N_{\text{pol}}$ (see table \ref{tab:resolutions}).

We see that as $L_{y}$ is increased the number of $y$ gridpoints must be increased proportionally. Additionally, after $L_{y}$ reaches a certain value, equation \refEq{eq:aspectRatio} forces us to increase $L_{x}$ (and thus the number of $x$ gridpoints) proportionally as well. However, for the parameters of tables \ref{tab:resolutions} and \ref{tab:CBC}, this was not necessary until $L_{y} > 1000 \rho_{i}$. For simulations with lower values of $\hat{s}$, for which parallel self-interaction is expected to be more of a concern, this constraint would kick in at lower values of $L_{y}$. On the other hand, when increasing $N_{\text{pol}}$ we clearly must increase the number of parallel gridpoints. Moreover, as explained in detail in \citet{WatanabeFluxTubeTrain2015}, properly resolving a longer parallel domain turns out to require more radial gridpoints, due to the fact that longer domains are twisted more by magnetic shear. Adding more radial modes decreases the timestep of explicit codes like GENE because we are allowing smaller radial scales, which makes the CFL limit more constraining. In light of this, it is not surprising that simulations with $N_{\text{pol}} > 1$ are so costly --- you must increase the parallel, radial, and temporal resolutions proportionally with $N_{\text{pol}}$. Additionally, pushing to extreme values of $L_{y}$ has the added advantage of ensuring that self-interaction in the $y$ direction is completely eliminated.

While increasing $L_{y}$ was the better way to reach convergence in our simulations, there are a few important caveats that make generalization difficult. First, at lower values of $\hat{s}$ the domain aspect ratio quantization condition (i.e. equation \refEq{eq:aspectRatio}) kicks in at lower values of $L_{y}$, making the $L_{y}$ route more costly. Second, codes that partially use implicit methods to advance in time may observe different computational costs than what we have seen in GENE, which uses only explicit methods, as they are not limited by the CFL condition. Third, adding perpendicular gridpoints (specifically when increasing $L_{x}$ and $L_{y}$ at fixed resolution) is fundamentally different than adding parallel gridpoints because the perpendicular modes are all coupled through a Fourier transform in the nonlinear term, while only some of the parallel gridpoints are connected via the parallel streaming term. Thus, the relative computational and communication costs may depend strongly on the computer architecture, memory, and parallelization scheme. Finally, \citet{WatanabeFluxTubeTrain2015} and \citet{ScottShiftedMetric2001} each present a clever way of potentially getting around the need to increase the radial and temporal resolutions with $N_{\text{pol}}$. While neither have been implemented in GENE, they both could make $N_{\text{pol}} > 1$ simulations much more affordable, potentially more affordable than large $L_{y}$ simulations. The first approach, the ``flux tube train'' \citep{WatanabeFluxTubeTrain2015}, has been implemented in the local gyrokinetic codes GKV \citep{WatanabeGKV2005} and \texttt{stella} \citep{BarnesStella2019}. The other approach, the ``shifted metric'' coordinate system \citep{ScottShiftedMetric2001}, has not been implemented in any local gyrokinetic code and it is currently unclear if it is possible to implement in a radially-periodic domain \citep{ToldShiftedMetric2010}.

As discussed in section \ref{sec:parallelBC}, the parallel self-interaction occurring in local simulations {\it is} physical when $L_{y}$ corresponds to a full flux surface. The same is true of self-interaction in the $y$ direction --- if you are modeling a full flux surface, the toroidal periodicity of a real device will lead to self-interaction if the toroidal correlation length is comparable to the toroidal circumference. Thus, you can also view the $N_{\text{pol}} = 1$ points in figure \ref{fig:resStudy} as a scan showing how the physical effect of self-interaction weakens with decreasing $\rho_{\ast}$\footnote{This convergence looks quantitatively similar to the system size investigations in \citet{McMillanSystemSizeScaling2010} and \citet{LinSystemSizeScaling2012}.}. This is indicated on the top horizontal axes in figure \ref{fig:resStudy}. For context, the DIII-D tokamak \citep{LuxonDIIID2002} has $\rho_{\ast} \approx 1 / 300$ and a full flux surface simulation corresponds to
\begin{align}
L_{y} = \frac{2 \pi}{q_{0}} \frac{1}{\rho_{\ast}} \frac{x_{0}}{a} \rho_{i} \approx 700 \rho_{i} . \label{eq:physicalLy}
\end{align}
Therefore, figure \ref{fig:resStudy} shows that self-interaction is a finite $\rho_{\ast}$ effect that we are usually justified in eliminating when modeling present-day large tokamaks. However, it might still play an important role in smaller machines or specific parameter regimes \footnote{Note that there are many other finite $\rho_{\ast}$ effects apart from the self-interaction discussed here. Some of these are contained in current global gyrokinetic codes (e.g. profile shearing), while many others have only been derived recently \citep{ParraLagrangianGyro2011} and are not in any code. Thus, while going to a full flux surface simulation will allow you to properly treat self-interaction, it is no guarantee that your overall simulation will be accurate for a small machines. On the contrary, the fact that self-interaction is important suggests that many of the other finite $\rho_{\ast}$ effects may be important too.}.

\begin{figure}
	\hspace{3em} (a)
	
	\begin{center}
		\includegraphics[width=0.7\textwidth]{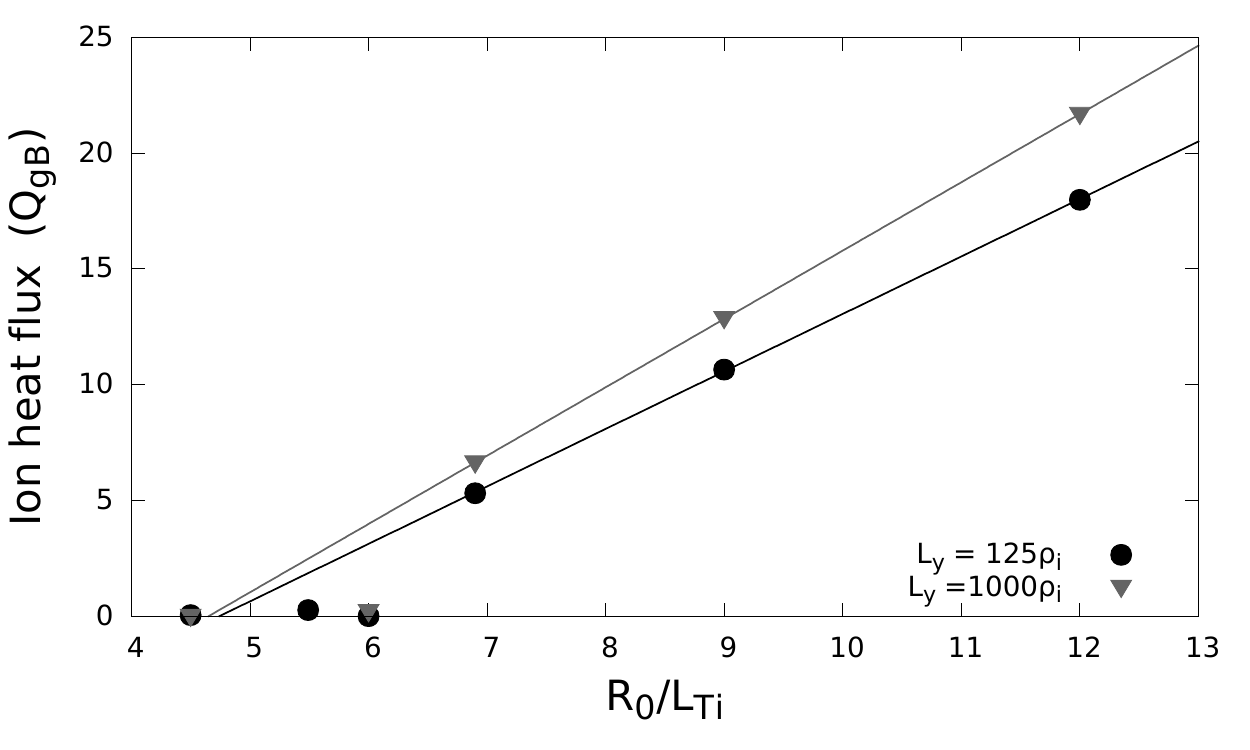}
	\end{center}
	
	\hspace{3em} (b)
	
	\begin{center}
		\includegraphics[width=0.7\textwidth]{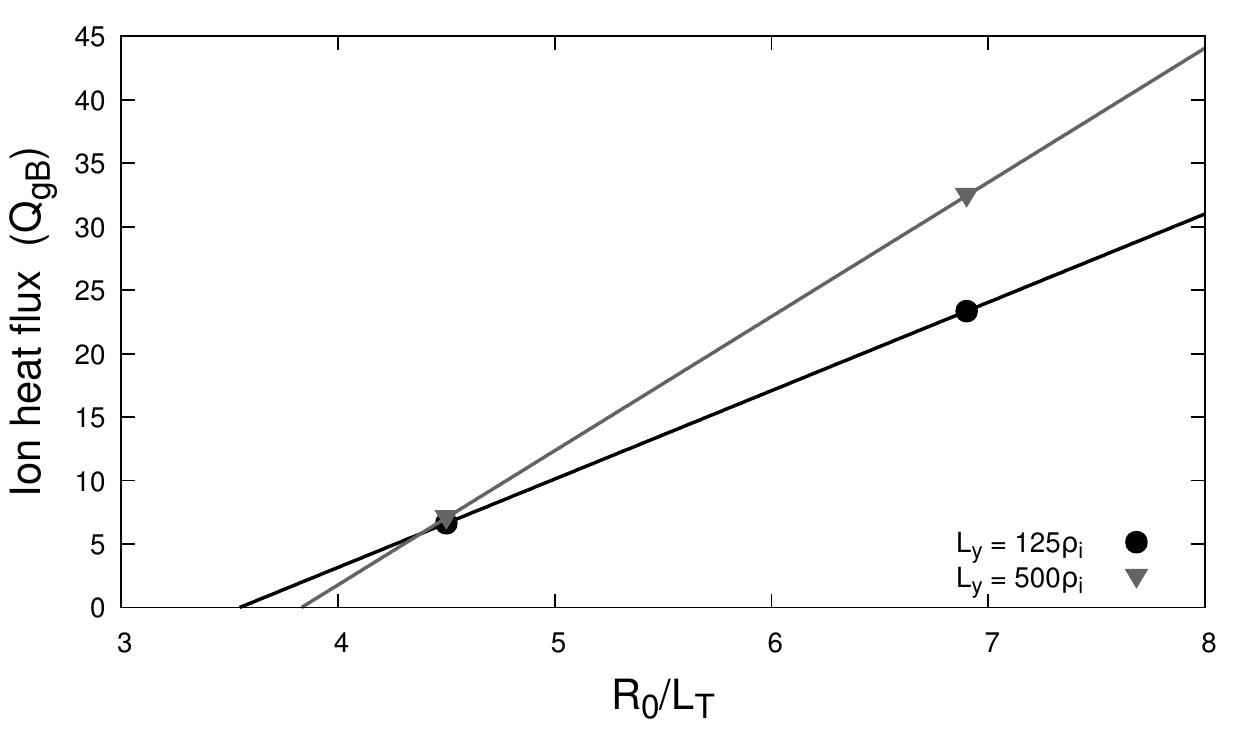}
	\end{center}
	\caption{A critical gradient study using adiabatic electrons (a) or kinetic electrons (b) for a domain width that mostly eliminates self-interaction (gray triangles) and another for a domain width that allows significant self-interaction (black circles). All simulations use the parameters of tables \ref{tab:resolutions} and \ref{tab:CBC} and best-fit linear trendlines are also shown.}
	\label{fig:critGrad}
\end{figure}

Lastly, we performed a temperature gradient scan to determine how self-interaction affects the critical gradient and profile stiffness in Cyclone Base Case simulations. Figure \ref{fig:critGrad} shows two data sets --- one with a large value of $L_{y}$, where self-interaction has been mostly diluted away, and one with a small value of $L_{y}$, where self-interaction is still significant. We see that the difference in the ion heat flux between the two values of $L_{y}$ is generally reduced as the temperature gradient decreases. This leads to similar estimates of the critical gradient for both sets of simulations, giving evidence that large $L_{y}$ domain widths may not be necessary for simulations near the critical gradient. However, the self-interaction present in the simulations with smaller $L_{y}$ reduces the profile stiffness, more significantly when using kinetic electrons.

\section{Conclusions}
\label{sec:conclusions}

In this work we have shown that turbulent self-interaction through the parallel boundary condition can significantly affect the results of local gyrokinetic simulations. Such self-interaction {\it can} be physical, but only when the simulation domain corresponds to a full flux surface for the device being modeled. Using a narrow flux tube to model a large device will artificially strengthen self-interaction, which can reduce the heat flux. This implies that, if the parallel self-interaction is significant, the simulation will only properly model it for a single finite value of $\rho_{\ast}$. However, if self-interaction is negligible, the simulation is appropriate for any sufficiently small value of $\rho_{\ast}$. In other words, self-interaction is a finite $\rho_{\ast}$ effect.

Thus, to achieve the true $\rho_{\ast} \ll 1$ limit assumed in deriving gyrokinetics, one would like to completely eliminate parallel self-interaction. We have shown in figure \ref{fig:resStudy} that this can be achieved by increasing the width of the simulation domain in the binormal direction $L_{y}$ and/or lengthening the simulation domain using $N_{\text{pol}}$. Currently available results suggest that self-interaction is stronger for kinetic electrons, low values of magnetic shear, and strong turbulence drive (i.e. steep background gradients). This may be because these parameters tend to increase the parallel correlation length of the turbulence. To verify that self-interaction has been eliminated, one should pay special attention to convergence in the binormal domain width and check for spikes in the parallel correlations function (e.g. figure \ref{fig:parCorr}). Additionally, one should take care when using a resolution study done with adiabatic electrons to justify domain resolutions for simulations using kinetic electrons.


The authors would like to thank B. McMillan, M. Hardman, B. Faber, A. Peeters, F. Parra, and M. Barnes for useful discussions pertaining to this work. 
This work was supported by the EUROfusion - Theory and Advanced Simulation Coordination (E-TASC). This work has been carried out within the framework of the EUROfusion Consortium and has received funding from the Euratom research and training programme 2014-2018 and 2019-2020 under grant agreement No 633053. The views and opinions expressed herein do not necessarily reflect those of the European Commission. 
We acknowledge the CINECA award under the ISCRA initiative, for the availability of high performance computing resources and support. 
Lastly, this work was supported by a grant from the Swiss National Supercomputing Centre (CSCS) under project ID s863 and s956.

\appendix
\section{Derivation of aspect ratio quantization}
\label{app:aspectRatio}

To see why the aspect ratio $L_{x}/L_{y}$ of a flux tube is quantized, we must start with the boundary conditions. The binormal boundary condition is
\begin{align}
  \phi ( x, y + L_{y}, \chi ) = \phi ( x, y, \chi ) , \label{eq:binormalBC}
\end{align}
where $y$ is defined by equation \refEq{eq:yDef}. Here we give the turbulent electrostatic potential $\phi$ as an example, but the same boundary conditions also hold for the electromagnetic perturbed fields as well as the gyrokinetic distribution function. The radial boundary condition is
\begin{align}
  \phi ( x + L_{x}, y, \chi ) = \phi ( x, y, \chi ) , \label{eq:radialBC}
\end{align}
where it is important that we are holding $y$ constant in applying radial periodicity. Lastly, the parallel condition is more complicated and is called the ``twist-and-shift''  \citep{BeerBallooingCoordinates1995}. It is defined by taking parallel periodicity at constant toroidal angle $\zeta$, according to
\begin{align}
  \phi ( x, y ( x, \zeta, \chi + 2 \pi N_{\text{pol}} ), \chi + 2 \pi N_{\text{pol}} ) = \phi ( x, y ( x, \zeta, \chi ), \chi ) . \label{eq:parallelBCgeneral}
\end{align}
This relation is also depicted in figure \ref{fig:boundaryGeo}. Using the definition of $y$ given by equation \refEq{eq:yDef} and linearizing the safety factor profile, we can see that
\begin{align}
  y ( x, \zeta, \chi + 2 \pi N_{\text{pol}} ) = y ( x, \zeta, \chi ) + 2 \pi N_{\text{pol}} x_{0} + 2 \pi N_{\text{pol}} \hat{s} \left( x - x_{0} \right) .
\end{align}
Thus, equation \refEq{eq:parallelBCgeneral} can be written as
\begin{align}
\phi ( x, y + 2 \pi N_{\text{pol}} \hat{s} \left( x - x_{0} \right), \chi + 2 \pi N_{\text{pol}} ) = \phi ( x, y, \chi ) , \label{eq:twistAndShiftBC}
\end{align}
where we have assumed that $2 \pi N_{\text{pol}} x_{0} / L_{y}$ is very close to an integer $N_{q} \in \mathbb{Z}$ and then applied the binormal periodicity (i.e. equation \refEq{eq:binormalBC}) $N_{q}$ times. This is valid because the minor radial location of the center of the flux tube $x_{0}$ is much larger than the binormal width $L_{y}$, so a slight radial shift ensures that the assumption is satisfied and doesn't affect the simulation.

The flux tube aspect ratio quantization condition fundamentally arises from the combination of the parallel and radial boundary conditions. As explained in the main text, if field lines are not connected properly across these two boundaries, parallel motion can lead to an artificial jump in the binormal direction. We begin by writing the radial boundary condition given by equation \refEq{eq:radialBC}. Then, replace both sides of the equation using the parallel boundary condition given by equation \refEq{eq:twistAndShiftBC} to find
\begin{align}
  \phi ( x +L_{x}, &y + 2 \pi N_{\text{pol}} \hat{s} \left( x + L_{x} - x_{0} \right), \chi + 2 \pi N_{\text{pol}} ) \label{eq:aspectRatioDeriv} \\
  &= \phi ( x, y + 2 \pi N_{\text{pol}} \hat{s} \left( x - x_{0} \right), \chi + 2 \pi N_{\text{pol}} ) .  \nonumber
\end{align}
Next, we again use the radial boundary condition given by equation \refEq{eq:radialBC} to replace the right side of equation \refEq{eq:aspectRatioDeriv} and show
\begin{align}
\phi ( x +L_{x}, &y + 2 \pi N_{\text{pol}} \hat{s} \left( x + L_{x} - x_{0} \right), \chi + 2 \pi N_{\text{pol}} ) \\
&= \phi ( x + L_{x}, y + 2 \pi N_{\text{pol}} \hat{s} \left( x - x_{0} \right), \chi + 2 \pi N_{\text{pol}} ) .  \nonumber
\end{align}
This expression can only be true if the $2 \pi N_{\text{pol}} \hat{s} L_{x}$ term in the second argument on the left side can be removed using repeated application of binormal periodicity (given by equation \refEq{eq:binormalBC}). Thus, we require that
\begin{align}
2 \pi N_{\text{pol}} \left| \hat{s} \right| L_{x} = N_{\text{si}} L_{y} \label{eq:condBinormalPeriodicity}
\end{align}
for some integer $N_{\text{si}} \in \mathbb{Z}$. This is the flux tube aspect ratio quantization condition given by equation \refEq{eq:aspectRatio}. If it is not satisfied, field line identity will not be consistent across the parallel boundary of the flux tube and spurious cross-field transport will be introduced into the simulation. A particle traveling purely along a field line can find itself on a different field line after passing through the parallel boundary, thereby enabling fictitious correlations between distant regions within the flux tube. Note that this is a separate issue from fictitious correlations that can be introduced by pseudo-integer surfaces (i.e. the primary focus of this paper). By applying equations \refEq{eq:toroidalShift} and \refEq{eq:binormalWidths} to equation \refEq{eq:condBinormalPeriodicity}, we can derive equation \refEq{eq:zetaShift} and show that the radial domain width must correspond to an integer number of lowest order pseudo-rational surfaces.

\bibliographystyle{jpp}

\bibliography{references}

\end{document}